\documentclass[twocolumn,pra,twocolumn,superscriptaddress]{revtex4}
\usepackage{color}
\usepackage{amsmath}
\usepackage{amssymb}
\usepackage{graphicx}
\usepackage{tikz}
\usepackage{multirow}

\usepackage{float}
\usepackage{bbm}
\usepackage{dsfont}

\usepackage{epsfig}
\usepackage{verbatim}
\usepackage{array}
\usepackage{setspace}

\usepackage[unicode=true,
bookmarks=true,bookmarksnumbered=false,bookmarksopen=false,
breaklinks=false,pdfborder={0 0 1},backref=false,colorlinks=true]
{hyperref}
\hypersetup{
	linkcolor=magenta, urlcolor=blue, citecolor=blue, pdfstartview={FitH}, hyperfootnotes=false, unicode=true}

\makeatletter
\@ifundefined{textcolor}{}
{%
	\definecolor{BLACK}{gray}{0}
	\definecolor{WHITE}{gray}{1}
	\definecolor{RED}{rgb}{1,0,0}
	\definecolor{GREEN}{rgb}{0,1,0}
	\definecolor{BLUE}{rgb}{0,0,1}
	\definecolor{CYAN}{cmyk}{1,0,0,0}
	\definecolor{MAGENTA}{cmyk}{0,1,0,0}
	\definecolor{YELLOW}{cmyk}{0,0,1,0}
}

\usepackage{amsfonts}\usepackage{tabularx}\usepackage{dcolumn}\usepackage{bm}\usepackage{graphicx}\usepackage{epstopdf}

\hypersetup{urlcolor=blue}

\makeatother

\newcolumntype{C}[1]{>{\centering\arraybackslash$}p{#1}<{$}}

\begin{document}
	
	\title{Sign-switching of superexchange mediated by few electrons under non-uniform magnetic field}
	
	\author{Guo Xuan Chan}
	\affiliation{Department of Physics, City University of Hong Kong, Tat Chee Avenue, Kowloon, Hong Kong SAR, China, and City University of Hong Kong Shenzhen Research Institute, Shenzhen, Guangdong 518057, China}
	
	\author{Xin Wang}
	\email{x.wang@cityu.edu.hk}
	\affiliation{Department of Physics, City University of Hong Kong, Tat Chee Avenue, Kowloon, Hong Kong SAR, China, and City University of Hong Kong Shenzhen Research Institute, Shenzhen, Guangdong 518057, China}
	\date{\today}
	
	\begin{abstract}
		Long range interaction between distant spins is an important building block for the realization of large quantum-dot network in which couplings between pairs of spins can be selectively addressed. Recent experiments on coherent logical states oscillation between remote spins facilitated by intermediate electron states has paved the first step for large scale quantum information processing. Reaching this ultimate goal requires extensive studies on the superexchange interaction on different quantum-dot spatial arrangements and electron configurations. Here, we consider a linear triple-quantum-dot with two anti-parallel spins in the outer dots forming the logical states while various number of electrons in the middle dot forming a mediator, which facilitates the superexchange interaction. We show that the superexchange is enhanced when the number of mediating electrons increases. In addition, we show that forming a four-electron triplet in the mediator dot further enhance the superexchange strength. Our work can be a guide to scale up the quantum-dot array with controllable and dense connectivity. 
	\end{abstract}
	\maketitle
	
	\section{Introduction}
\begin{figure}[t]
		\includegraphics[width=\linewidth]{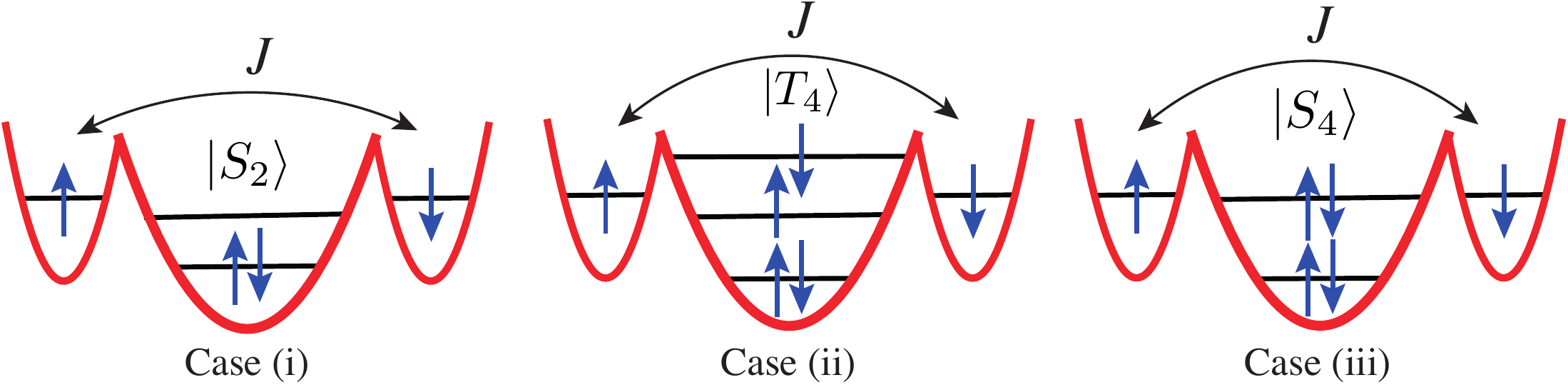}
		\caption{Three different electron configurations studied in this work. Case (i): Two electrons in the mediator (middle dot) form a singlet ($S=0$), where $S$ is the total spin. Case (ii): Four electrons in the mediator form a triplet ($S=1$). Case (iii): Four electrons in the mediator form a singlet ($S=0$). $J$ is the superexchange energy between distant spins in the outer-dots.}
		\label{fig:threecase}
	\end{figure}

	Early studies of semiconductor quantum-dot qubits, both experimental \cite{Levy.02,Hayashi.03,Petta.05,Gorman.05,Shinkai.09,Petersson.10,Laird.10,Barthel.10,Shi.11,Dovzhenko.11,Maune.12,Shulman.12,Shi.13,Wu.14,Eng.15,Reed.16,Martins.16,Harvey.17,Malinowski.17,Noiri.18,Takeda.20,Cerfontaine.20} and theoretical \cite{Medford.13,Taylor.13,Yang.17,Yang.18,Shim.18,Abadillo.19,Xie.21,Friesen.17,Sala.17,Zhang.18}, mostly focus on interactions between neighboring electron spins. Approaches to harness non-proximal exchange interaction between distant spins is critical to achieve efficient non-local quantum operations, as increased connectivity leads to smaller quantum circuit depth \cite{Fei.12}. Current implementations of long range interaction between spins includes capacitive coupling \cite{Shulman.12,Li.15,Nichol.17,Neyens.19}, photon mediated interactions \cite{Mi.17,Mi.17.2,Mi.18,Woerkom.18,Koski.20,Borjans.20,Burkard.20,Borjans.20.2,Kratochwil.21} and electron shuttling \cite{Nakajima.18,Fujita.17,Jadot.21,Mills.19,Buonacorsi.20,Ginzel.20}. The former two schemes introduce coupling to the charge degree of freedom, hence they are prone to decoherence by charge noise \cite{Shulman.12,Nichol.17,Koski.20} while the later method requires a relatively complex operation and coordination of the gate voltages to perform electron shuttling adiabatically \cite{Mills.19}. An alternative method to have exchange interactions between remote spins is enabling a virtual exchange through a mediator, termed as superexchange \cite{Chan.21,Fei.12,Qiao.21,Malinowski.19,Baart.17,Qiao.20,Braakman.13}. Current experimental progress has demonstrated coherent superexchange interaction with the mediator being an empty \cite{Baart.17}, singly occupied \cite{Chan.21} or multi-electron \cite{Malinowski.19} quantum-dot. Among different superexchange coupling schemes, implementing a multi-electron quantum-dot or dot-chain is of interest as it has been shown theoretically that larger number of the electrons occupying the mediator leads to stronger superexchange interaction \cite{Deng.20}. In addition, current works mostly focus on a spinless multi-electron mediator \cite{Deng.20,Malinowski.19,Qiao.20,Qiao.21}, leaving the effect of non-zero spin state formed in the mediator or larger number of electrons occupying the mediator unanswered. In this manuscript, we explore, using Configuration Interaction (CI) calculations, the effect on the superexchange in a linear triple-quantum-dot (TQD) system with the mediator occupied by two or four electrons. Specifically, we compare three different cases for the mediator (summarized in Fig.~\ref{fig:threecase}): (i) a spinless two-electron state, (ii) a spin-1 four-electron state and (iii) a spinless four-electron state. 
	We have found that 
	in contrast to the spinless two-electron mediator, which yields positive superexchange energy, spinless four-electron mediator results in negative superexchange, with stronger magnitude. Furthermore, if larger perpendicular magnetic field is applied on the outer two dots in a TQD device as compared to the inner dot, for a spin-1 four-electron mediator, the superexchange, denoted as $J$, is negative for moderate magnetic field while switches to positive value for much larger magnetic field, with $J$ in the the former case yielding larger magnitude than the later. 
	
	The paper is organized as follows: In Sec.~\ref{sec:model}, the model and methods are provided. In Sec.~\ref{subsec:singleMultielectron}, we evaluate the spin of four-electron mediator under different perpendicular magnetic field strengths. In Sec.~\ref{subsubsec:superexchangeSixT} and  \ref{subsubsec:superexchangeSixS}, we study the superexchange mediated by a four-electron triplet and singlet respectively. In Sec.~\ref{subsubsec:compare}, we compare the results presented in Sec.~\ref{subsubsec:superexchangeSixT} and  Sec.~\ref{subsubsec:superexchangeSixS}, along with the superexchange mediated by a two-electron singlet. In the end, we summarize our results in Sec.~\ref{sec:conclusion}. An appendix is provided to give more details on CI calculations (Appendices \ref{sec:conv} and \ref{sec:Grd2es}), the four-electron triplet mediator (Appendix \ref{sec:4esTriplet}) and the leakage estimation for the superexchange mediated by a triplet state (Appendix \ref{sec:lea}).
	
	\section{Model and methods}\label{sec:model}
\begin{figure}[t]
		\includegraphics[width=0.98\linewidth]{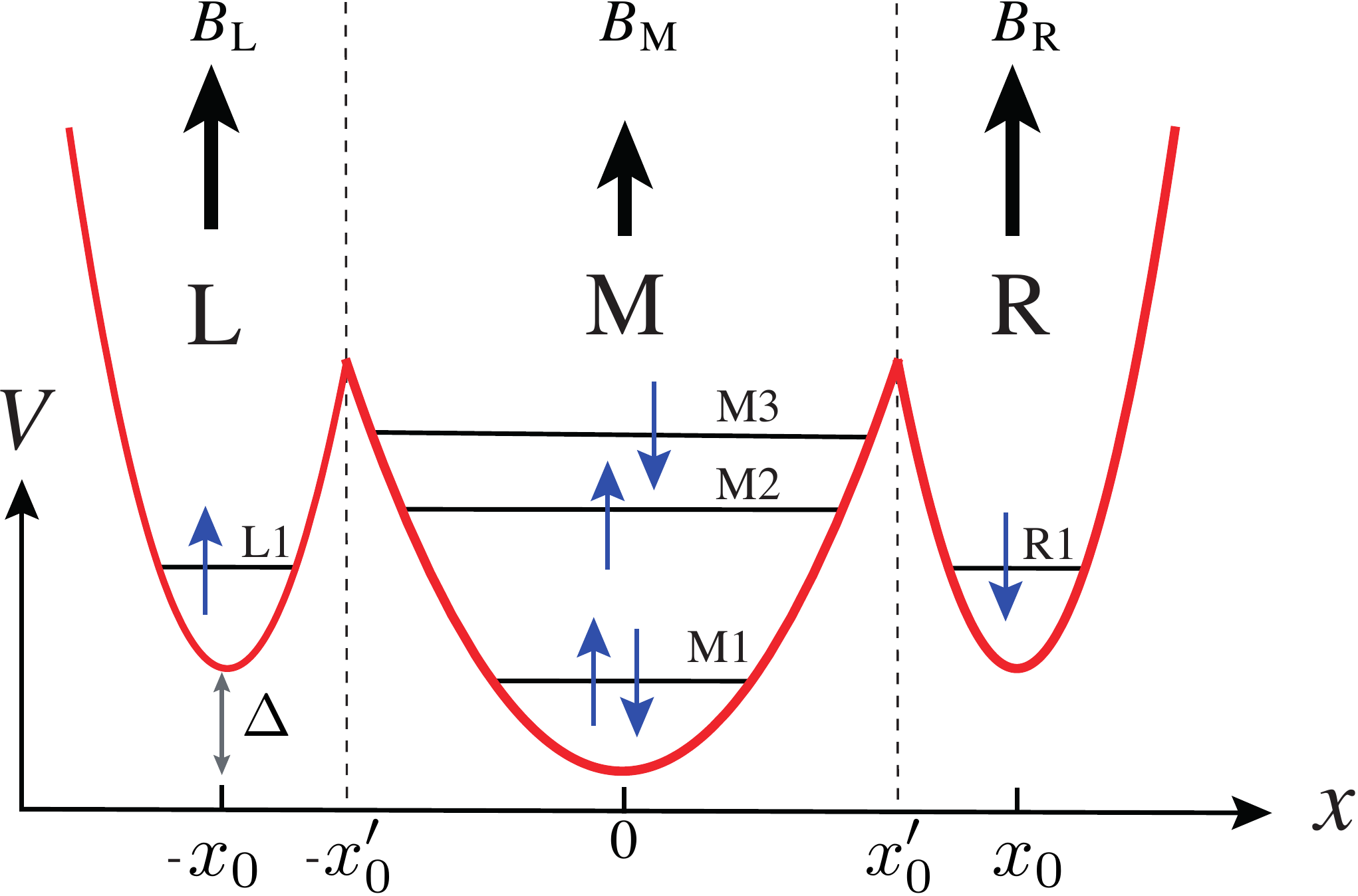}
		\caption{Schematic illustration of the model potential given in Eq.~\eqref{eq:V}. The dashed lines mark the boundaries between adjacent potential wells.}
		\label{fig:potential}
	\end{figure}

	We consider an $N$-electron system described by the Hamiltonian
	
\begin{equation}	
H=\sum_{j=1}^N h_j + \sum_{j<k}{e^2}/\epsilon\left|\mathbf{r}_j-\mathbf{r}_k\right|,
\end{equation}
with the single-particle Hamiltonian $h_j$ being
\begin{equation}
h_j = {(-i\hbar \nabla_j+e \mathbf{A}_j/c)^2}/{2m^*}+V(\mathbf{r}_j)+g^*\mu_B \mathbf{B}_j\cdot \mathbf{S}_j,
\end{equation}
where $\mathbf{r}_j=x_j \widehat{\mathbf{x}}+ y_j \widehat{\mathbf{y}}$ indicates the position of the $j$th electron with spin $\mathbf{S}_j$ and experiencing a perpendicular magnetic field $\mathbf{B}_j=B_j \widehat{\mathbf{z}}$. $\mathbf{A}_j$ is the vector potential corresponding to $\mathbf{B}_j$ and  $m^*$ is the effective mass, taken as 0.067 electron mass in GaAs. In this work, $V(\mathbf{r})$ describes the confinement potential of a triple-quantum-dot system (TQD), modeled as (cf.~Fig.~\ref{fig:potential}): 
	\begin{equation}\label{eq:V}
		V(\mathbf{r})=
		\begin{cases}
			\frac{1}{2}m^*\omega_\text{L}^2\left(\mathbf{r}-\mathbf{R}_\text{L}\right)^2 & x < -x'_0,\\
			\frac{1}{2}m^*\omega_\text{M}^2\left(\mathbf{r}-\mathbf{R}_\text{M}\right)^2+\Delta & -x'_0<x<x'_0,\\
			\frac{1}{2}m^*\omega_\text{R}^2 \left(\mathbf{r}-\mathbf{R}_\text{R}\right)^2 & x>x'_0.
		\end{cases}
	\end{equation}
Here, $\mathbf{R}_\text{L} = \left(-x_0,0\right)$,  $\mathbf{R}_\text{M} = \left(0,0\right)$, and $\mathbf{R}_\text{R} = \left(x_0,0\right)$ are the minima of the three parabolic wells, representing three dots labeled as L (left), M (middle) and R (right), respectively. $\omega_\text{L}$, $\omega_\text{M}$, $\omega_\text{R}$ and $B_\text{L}$, $B_\text{M}$, $B_\text{R}$ are the confinement strengths and the magnetic field at dots L, M, R respectively. The confinement strengths are set as $\hbar \omega_\text{L} = \hbar \omega_\text{R} = 7.28$meV, $\hbar \omega_\text{M} = $ 3.64meV \cite{Deng.20,Nielsen.13}, while the magnetic field varies as will be explained in the following sections. The potential cuts, $- x'_0$ and $x'_0$, are determined by locating the values of $x$ at which the potential values of dots L and  M, and dots M and R are equal, respectively, at $y=0$.

	We use the CI method to solve the multielectron problem, which involves using orthonormalized Fock-Darwin states (F-D) to approximate the single electron wavefunctions in a quantum-dot. The orthonormalized F-D states are obtained by Cholesky decomposition of the overlap matrix formed by the bare F-D states \cite{Barnes.11}. A rigorous description of the multielectrom eigenstates requires keeping all the F-D states, which is forbiddingly expensive. In practice, one truncates the number of F-D states in each quantum-dot while maintaining the convergence. Here, we use a cut-off scheme to keep the CI calculation tractable following Ref.~\onlinecite{Barnes.11}. In this scheme, only the multielectron Slater determinants whose non-interacting energies are within the predefined cut-off values are retained. The cut-off values are defined to be the maximum achievable energy by a Slater determinant with one electron occupying the highest F-D state while the remaining electrons occupying the ground orbital. We keep 10 lowest F-D states (corresponding to the principal quantum number $n$ up to 3) in each dot as suggested by the convergence of the ground energy of four-electron state in a quantum dot, cf.~Fig.~\ref{fig:GrdStEvalSingleDot}, and the convergence of the exchange energy of two-electron state in an undetuned double-quantum-dot (DQD) device, cf.~Fig.~\ref{fig:twoElectronDQDExchange}. The calculations for the six-electron system are carried out on a high-performance computing cluster with 480 2.2GHz Intel Xeon CPUs and 42GB of memory, where a data point at a given detuning, $\Delta$, typically costs 40 minutes.
	\section{Results} \label{sec:resut}
	\subsection{Multielectron single dot} \label{subsec:singleMultielectron}
\begin{figure}[t]
		\centering
		\includegraphics[width=0.98\columnwidth]{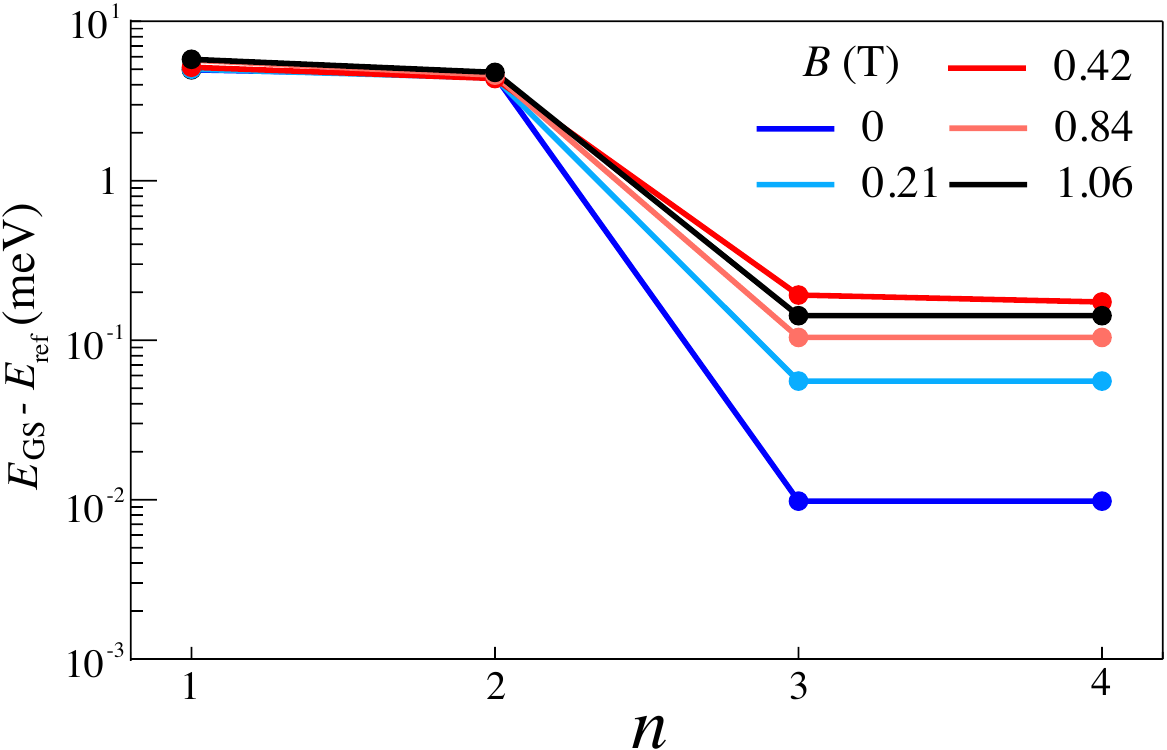}
		\caption{Ground state energy of a quantum-dot occupied by four electrons versus the maximum principal quantum number $n$ of orbitals kept in the calculation. $n=1$ corresponds to a total of 3 orbitals, while $n=2, 3, 4$ correspond to $6, 10, 15$ orbitals respectively. The parameters are: $\hbar \omega_0 = 3.64$ meV, $E_\text{ref} = $ 48.06meV.}
		\label{fig:GrdStEvalSingleDot}
	\end{figure}

	Experiments have shown that a multielectron quantum-dot system at zero magnetic field may exhibit negative exchange interaction, which means that the (unpolarized) triplet state \cite{triplet} yields lower energy than the singlet state. In these works, the total number of electrons ranges from as low as four \cite{Deng.18,Kalliakos.08,Kouwenhoven.97} to as many as 50-100 \cite{Martins.17,Malinowski.18}. To facilitate the theoretical discussion, we focus on a quantum-dot system occupied by four electrons. We vary the strength of the magnetic field and have found that 
	the ground state switches from the triplet state at weak field to the singlet state at sufficiently large magnetic field, cf.~Table~\ref{tab:groundStateOfsingleDot} \cite{ChanGX.21}. It is also known that the ground state in a similar system can be the singlet state when the eccentricity of the quantum dots is increased beyond certain threshold at zero magnetic field \cite{Deng.18}. In this paper, we focus on circular quantum dots only. The ground state energy of the corresponding four-electron states are shown in Fig.~\ref{fig:GrdStEvalSingleDot}.
		\begin{table}[!ht]
		\begin{tabular}{|p{1cm}|p{3cm}|p{3cm}|}
			\cline{1-3}
			\multirow{2}{*}{$B$ (T)} & \multicolumn{2}{c|}{Ground state configuration} \\
			\cline{2-3}
			&  \hfil 2e & \hfil 4e\\
			\cline{1-3}
			\hfil 0 & \hfil S & \hfil T \\
			\cline{1-3}
			\hfil 0.21  & \hfil S & \hfil T \\ 
			\cline{1-3}
			\hfil 0.42  & \hfil S & \hfil T \\
			\cline{1-3}
			\hfil 0.84  & \hfil S & \hfil S \\
			\cline{1-3}
			\hfil 1.06  & \hfil S & \hfil S \\
			\cline{1-3}
		\end{tabular}
		\caption{Ground state configuration of a quantum-dot occupied by two (2e) and four electrons (4e) based on the cut-off scheme with varying magnetic field. S indicates a singlet and T a triplet. The confinement energy of the quantum dot is $\hbar \omega_0 =  3.64$ meV. 10 orbitals (up to $n=3$) are retained for the single quantum-dot in the CI calculation.}
		\label{tab:groundStateOfsingleDot}
	\end{table}

	\subsection{Multielectron mediated superexchange, $J$} \label{subsec:superexchange}
	We denote the electron occupation in each dot as $(n_\text{L},n_\text{M},n_\text{R})$, where $n_\text{L}$ $(n_\text{M},n_\text{R})$ indicates the number of electrons in left  (middle, right) dot. For descriptive purpose, we denote the ground state configuration formed by $n_\text{M}$ electrons in dot M as $|S_{n_\text{M}}\rangle$ and $|T_{n_\text{M}}\rangle$ for the singlet and triplet states respectively. The dot parameters for different cases are summarized in Table~\ref{tab:summaryCompare}. The exchange energies are evaluated at the detuning, $\Delta$, such that the number of electrons in dot M is $n_\text{M}$.
	
	For a TQD device with six electrons in the (1,4,1) region, there exist two cases for the four-electron state in the middle dot. As suggested by the results based on a single dot (Table \ref{tab:groundStateOfsingleDot}), at weak magnetic fields ($B\lesssim 0.42$ T), we would expect the electrons to form a triplet ground state, $|T_4\rangle$. On the other hand, at larger magnetic fields ($B\gtrsim 0.84$ T), the electrons form a singlet ground state, $|S_4\rangle$. For illustrative purpose, throughout this manuscript, we plot the negative $J$ as dashed lines while positive $J$ as solid lines.
	\subsubsection{Six-electron TQD under weak magnetic field on the middle dot (Case (ii))} \label{subsubsec:superexchangeSixT}
	
	\begin{figure}[t]
		\includegraphics[width=0.98\columnwidth]{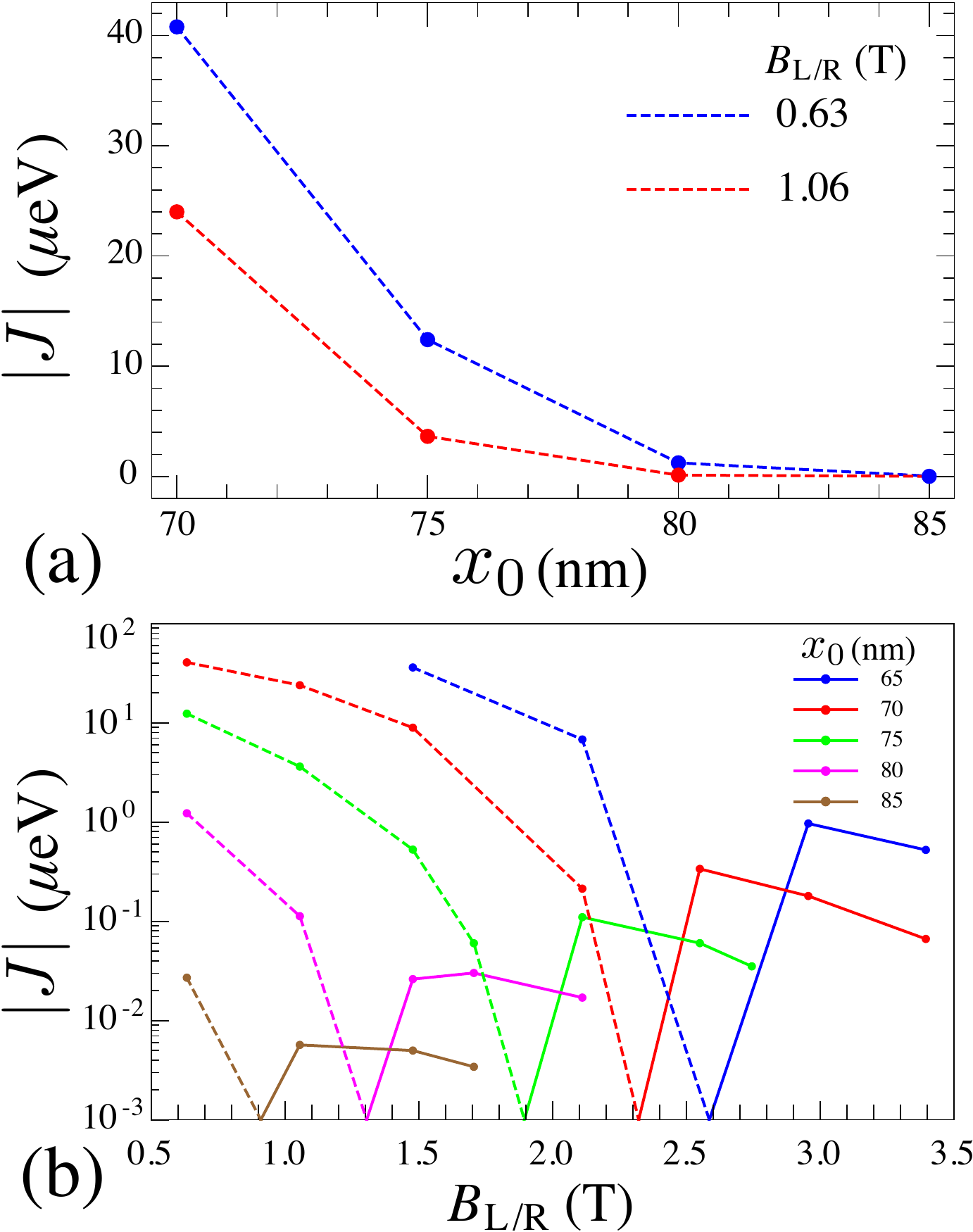}
		\caption{(a) Absolute value of the superexchange energy, $|J|$, as a function of interdot distance, $x_0$, for a six-electron system. (b) Exchange splitting, $J$, as a function of magnetic field applied on outer dots (L and R), $B_\text{L/R}$, for a six-electron system. Magnetic field of strength $B_\text{M}=0.21 $T is applied at dot M. Note that dashed lines represent negative values and solid lines positive values.}
		\label{fig:J4esvsx0BlrLarger}
	\end{figure}

	The low energy subspace of the six-electron system with $S_z = 0$ is spanned by six eigenstates, as provided in Table~\ref{tab:tripletGroundEigenstates}. At weak magnetic fields ($B\lesssim 0.42$ T), since four electrons in dot M form a triplet ground state, four of the six eigenstates with triplet ground state formed in dot M span the low-energy subspace. For a singlet-triplet qubit defined by anti-parallel spins in dots L and R, we denote the logical states as $|\widetilde{S}\rangle$ and $|\widetilde{T}\rangle$ for singlet and triplet states respectively, while the leakage states as $|\widetilde{T}_+\rangle$ and $|\widetilde{T}_-\rangle$, cf.~Eq.~\eqref{eq:logicalLeakageSt}. The superexchange energy, $J$, is defined as the energy difference between $|\widetilde{S}\rangle$ and $|\widetilde{T}\rangle$, which is evaluated at the detuning such that $n_\text{M}=4$. When the applied magnetic field is uniform across the TQD, the logical and leakage state is highly mixed, cf.~Table~\ref{tab:tripletGroundEigenstates}. The leakage can be suppressed by applying different magnetic fields on the outer and inner dots, i.e. $B_\text{L} = B_\text{R} \neq B_\text{M}$. We found that for  parameters relevant to this work, leakage into $|\widetilde{T}_\pm\rangle$ is smaller than $10^{-3}$ for $B_\text{L/R} - B_\text{M} \gtrsim 0.42$ T, cf.~Appendix~\ref{sec:lea} and Fig.~\ref{fig:leaVSBfoLR}. The corresponding $J$ as function of inter-dot distance is shown in Fig.~\ref{fig:J4esvsx0BlrLarger}(a). It is observed that $J$ is negative whose absolute value decreases when the inter-dot distance increases. Fig.~\ref{fig:J4esvsx0BlrLarger}(b) shows $J$ for $B_\text{L/R}$ ranging from $B_\text{L/R}=3B_\text{M}$ to $B_\text{L/R}\gg B_M$. It is observed that, for small $B_\text{L/R}$, $|J|$ decreases as $B_\text{L/R}$ is increased. Beyond certain threshold value of $B_\text{L/R}$, $J$ switches sign, becoming positive, and increases as a function of $B_\text{L/R}$. After it reaches a peak value, it decreases again but maintains its positive value.
		
	\subsubsection{Six electron-triple quantum-dot with larger magnetic field (Case (iii))}
	\label{subsubsec:superexchangeSixS}
	\begin{figure}[t]
		\includegraphics[width=0.9\columnwidth]{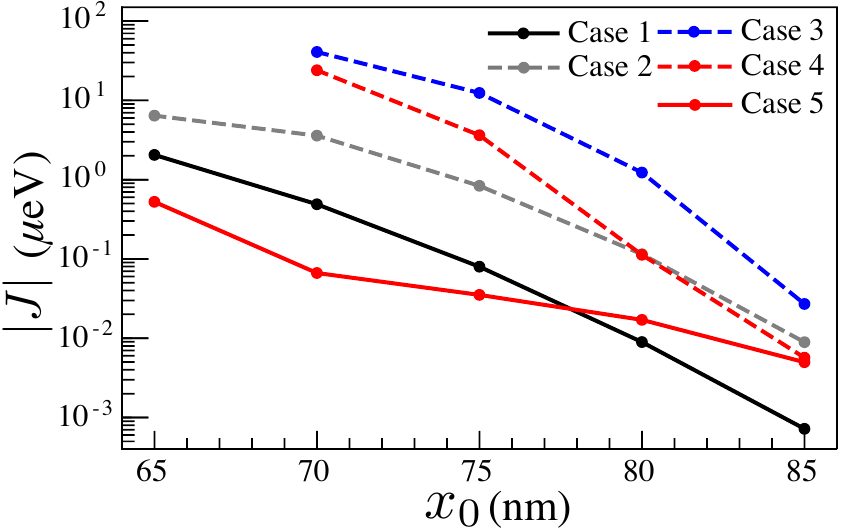}
		\caption{Absolute value of the superexchange energy, $|J|$, as a function of interdot distance, $x_0$, for a four-electron system (blue dashed line), a six-electron system in which MED forming singlet ground state (red dashed line) and triplet ground state (red solid line). The summary of different cases is provided in Table~\ref{tab:summaryCompare}.}
		\label{fig:exchangeEnergy}
	\end{figure}

	\begin{table}[t]
		\begin{tabular}{|c|c|c|c|}
			\cline{1-4}
			\multirow{2}{*}{Case} & \multirow{2}{*}{$(n_\text{L},n_\text{M},n_\text{R})$} & \multirow{2}{*}{$B_\text{L/M/R}$ (T)} & ground state \\
			& & & in dot M \\
			\cline{1-4}
			1 & $(1,2,1)$ & $B_\text{L/M/R} = 0.86$ & $|S_2\rangle$\\
			\cline{1-4}
			2 &\multirow{4}{*}{$(1,4,1)$} & $B_\text{L/M/R} = 0.86$ & $|S_4\rangle$ \\
			\cline{1-1}
			\cline{3-4}
			3& & $B_\text{M} = 0.21$, $B_\text{L/R} = 3 B_\text{M}$& \multirow{3}{*}{$|T_4\rangle$}\\
			\cline{1-1}
			\cline{3-3}
			4& & $B_\text{M} = 0.21$, $B_\text{L/R} = 5 B_\text{M}$& \\
			\cline{1-1}
			\cline{3-3}
			5& & $B_\text{L/R}  \gg B_\text{M}$& \\
			\cline{1-4}
		\end{tabular}
		\caption{Summary of different cases in Fig.~\ref{fig:exchangeEnergy}. $|S_{n_\text{M}}\rangle$ and $|T_{n_\text{M}}\rangle$ denote the singlet ground state and triplet formed by $n_\text{M}$ electrons in dot M respectively.}
		\label{tab:summaryCompare}
	\end{table}
	When a larger uniform magnetic field is applied across the TQD device, the lowest logical subspace is free of leakage since four electrons in dot M form a singlet ground state, $|S_4\rangle$. The logical states are shown explicitly in Table~\ref{tab:singletGroundEigenstates}. We observe that $J$ is negative when the mediating four-electron-state in dot M forms a singlet ground state, cf.~gray dashed line in Fig.~\ref{fig:exchangeEnergy}.
	\begin{table}[t]
		\begin{tabular}{|c|c|c|}
			\cline{1-3}
			& & \\[-9.5pt]
			Label &$S^2$ & Eigenstates \\
			\cline{1-3}
			& & \\[-10pt]
			$|S^\text{O}S^\text{I}\rangle$ &0 & $\left(|\uparrow_\text{R1}\downarrow_\text{L1}\rangle+|\uparrow_\text{L1}\downarrow_\text{R1}\rangle\right)|\uparrow_\text{M1}\downarrow_\text{M1}\uparrow_\text{M2}\downarrow_\text{M2}\rangle$ \\
			\cline{1-3}
			& & \\[-10pt]
			$|T^\text{O}S^\text{I}\rangle$ &2 & $\left(|\uparrow_\text{R1}\downarrow_\text{L1}\rangle-|\uparrow_\text{L1}\downarrow_\text{R1}\rangle\right)|\uparrow_\text{M1}\downarrow_\text{M1}\uparrow_\text{M2}\downarrow_\text{M2}\rangle$ \\
			\cline{1-3}
		\end{tabular}
		\caption{Eigenstates of the subspace formed by six electrons when four electrons in dot M form a triplet ground state.}
		\label{tab:singletGroundEigenstates}
	\end{table}
 	
 	\subsubsection{Comparison between two and four-electron mediated superexchange (Case (i), (ii) and (iii))}
 	\label{subsubsec:compare}
 	Figure~\ref{fig:exchangeEnergy} shows the superexchange, $J$, as a function of the inter-dot distance for several cases, including $J$ mediated by two electrons   (solid black line, Case (i)), $J$ mediated by four-electron singlet (dashed gray line, Case (ii)) and $J$ mediated by  four-electron triplet  (dashed blue, dashed red and solid red line for different $B_\text{L/R}$, Case (iii)). The data points of solid red line are extracted from Fig.~\ref{fig:J4esvsx0BlrLarger}(b) at the second values of $B_\text{L/R}$ which give positive $J$. 
 	
 	We first focus on the six-electron system. At smaller $x_0$, $J$ mediated by four-electron triplet ($\left|T_4\right\rangle$) yields larger magnitude than $J$ mediated by four-electron singlet ($\left|S_4\right\rangle$). In particular, $J$ mediated by $\left|T_4\right\rangle$ with $B_\text{L/R} = 3 B_\text{M}$ yields an absolute value that is around one order-of-magnitude larger than $J$ mediated by $\left|S_4\right\rangle$ (compare blue dashed line, Case 3, and gray dashed line, Case 2, for $x_0 \lesssim 75$nm in Fig.~\ref{fig:exchangeEnergy}). On the other hand, at large $x_0$ ($x_0 = 85$nm), $J$ mediated by $\left|S_4\right\rangle$ and $\left|T_4\right\rangle$ give comparable $|J|$.
 	
 	We next compare $J$ mediated by two-electron and four-electron states in dot M. Overall, four-electron mediated $J$ is stronger compared to two-electron ($|S_2\rangle$) mediated $J$, except for the case in which $B_\text{L,R}\gg B_M$. In particular, $\left|S_4\right\rangle$ mediated $J$ is around one order-of-magnitude stronger than $|S_2\rangle$ mediated $J$ while $\left|T_4\right\rangle$ mediated $J$ at $B_\text{L/R} = 3B_\text{M}$ is about two orders of magnitude stronger. Such magnetic gradient, $\Delta B = B_\text{L/R} - B_\text{M}= 0.42$ T $=10\mu eV=2.5\text{GHz}\times h$ should be achievable in near-term quantum devices \cite{Chesi.14}, since magnetic gradient as high as $\Delta B = 1\text{GHz}$ has been demonstrated \cite{Nichol.17}.
	It should be noted that larger $J$ can be achieved for $|S_2\rangle$ at much smaller $x_0$ \cite{Deng.20}. However, $|S_4\rangle$ or $|T_4\rangle$ is not achievable in that regime because for small inter-dot distance, the (1,4,1) dot occupation is not well defined.
	 	
 	\section{Conclusions}\label{sec:conclusion}
	We have shown, using full CI calculations, that in a TQD device, the variation of the number of electrons in the mediator has a considerable impact on the superexchange. We have observed that the magnitude of superexchange decreases as the inter-dot distances are increased. We have found that the four-electron mediator yields a stronger superexchange as compared to two-electron mediator for the same inter-dot distance in most cases. We have further shown that, in comparison to the four-electron singlet in the mediating dot, the four-electron triplet exhibits stronger superexchange, except when the magnetic field on the outer dots is much stronger than on the middle dot. Our results therefore should facilitate realization of large scale architecture with long range connectivity for quantum-dot spin qubit.
	
\section*{Acknowledgements} We acknowledge support from the Key-Area Research and Development Program of GuangDong Province  (Grant No.~2018B030326001), the National Natural Science Foundation of China (Grant No.~11874312), the Research Grants Council of Hong Kong (Grant No.~11303617), and the Guangdong Innovative and Entrepreneurial Research Team Program (Grant No.~2016ZT06D348). The calculations involved in this work are mostly performed on the Tianhe-2 supercomputer at the National Supercomputer Center in Guangzhou, China.

	\appendix
	\section{Singly occupied double-quantum-dot device}\label{sec:conv}
	\renewcommand{\theequation}{A-\arabic{equation}}
	\renewcommand{\thefigure}{A-\arabic{figure}}
	\setcounter{equation}{0}
	\setcounter{figure}{0}
	Figure~\ref{fig:twoElectronDQDExchange} shows the eigenvalues of the lowest singlet and triplet state of an undetuned double-quantum-dot (DQD) device occupied by two electrons. It can be observed that the exchange energy converges for $n \geq 3$.
	\begin{figure}[t]
		\centering
		\includegraphics[width=0.9\columnwidth]{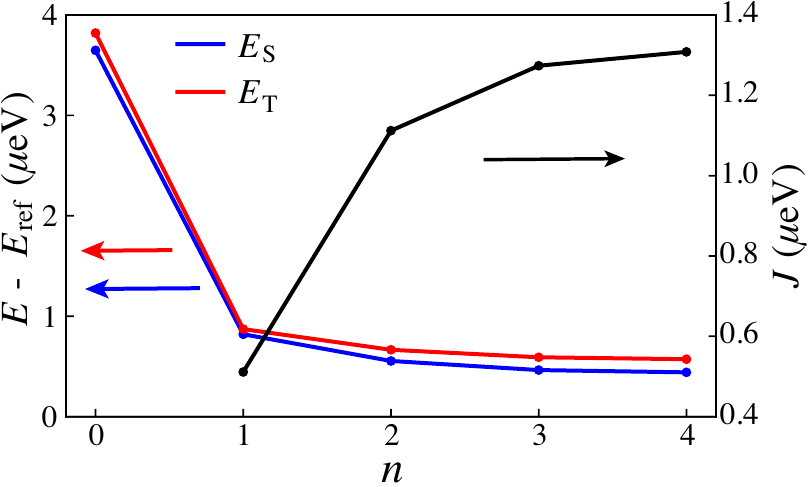}
		\caption{Lowest singlet and triplet eigenvalues and the corresponding exchange energy versus the maximum principal quantum number $n$ of orbitals kept in the CI calculation. The parameters are: inter-dot distance $x_0 = 28$ nm, $\hbar\omega_0$ = 7.28 meV, $B = 0.21$ T.}
		\label{fig:twoElectronDQDExchange}
	\end{figure}

	\section{Ground state energy of two-electron state occupying a quantum-dot }\label{sec:Grd2es}
	\renewcommand{\theequation}{B-\arabic{equation}}
	\renewcommand{\thefigure}{B-\arabic{figure}}
	\setcounter{equation}{0}
	\setcounter{figure}{0}
	Figure~\ref{fig:grdEval2es} shows the ground state energy when there are two electrons occupying a quantum  dot. It is observed that the ground state energy converges at $n=2$.
	\begin{figure}[t]
		\includegraphics[width=0.9\columnwidth]{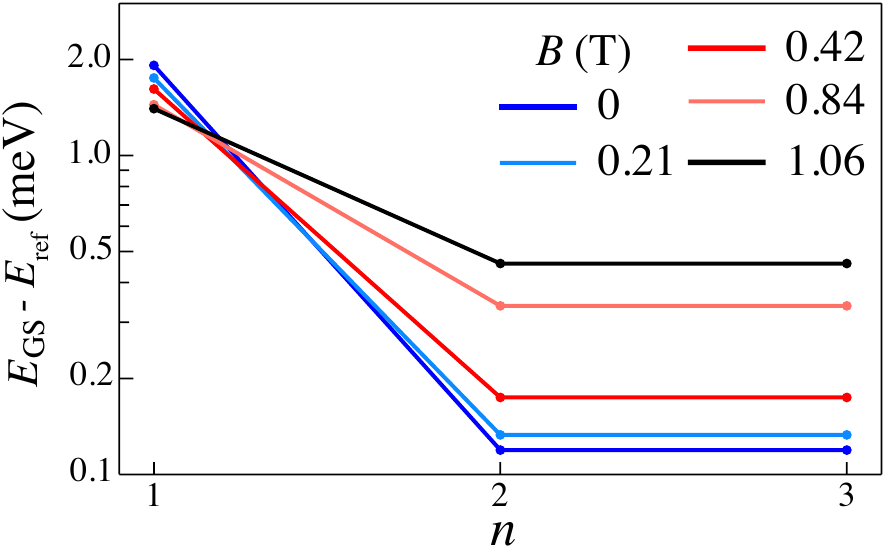}
		\caption{Ground state energy of a quantum-dot occupied by two electrons versus the maximum principal quantum number $n$ of orbitals kept in the calculation.}
		\label{fig:grdEval2es}
	\end{figure}
	
	\section{Eigenstates of six electron system with four electrons in the middle dot forming a triplet}\label{sec:4esTriplet}
	\renewcommand{\theequation}{C-\arabic{equation}}
	\renewcommand{\thefigure}{C-\arabic{figure}}
	\renewcommand{\thetable}{C-\arabic{table}}
	\setcounter{equation}{0}
	\setcounter{figure}{0}
	\setcounter{table}{0}
	\begin{table*}[t]
		\begin{tabular}{|c|c|c|}
			\cline{1-3}
			& & \\[-9.5pt]
			Label &$S^2$ & Eigenstates \\
			\cline{1-3}
			& & \\[-10pt]
			$|T^\text{O}T^\text{I}\rangle$&0 & $2|\uparrow_\text{M2}\uparrow_\text{M3}\downarrow_\text{R1} \downarrow_\text{L1}\rangle+2|\uparrow_\text{R1}\uparrow_\text{L1}\downarrow_\text{M2} \downarrow_\text{M3}\rangle-\left(|\uparrow_\text{R1}\downarrow_\text{L1}\rangle-|\uparrow_\text{L1}\downarrow_\text{R1}\rangle\right)\left(|\uparrow_\text{M2}\downarrow_\text{M3}\rangle-|\uparrow_\text{M3}\downarrow_\text{M2}\rangle\right)$ \\
			\cline{1-3}
			& & \\[-10pt]
			$|S^\text{O}S^\text{I}\rangle$ &0 & $\left(|\uparrow_\text{R1}\downarrow_\text{L1}\rangle+|\uparrow_\text{L1}\downarrow_\text{R1}\rangle\right)\left(|\uparrow_\text{M2}\downarrow_\text{M3}\rangle+|\uparrow_\text{M3}\downarrow_\text{M2}\rangle\right)$ \\
			\cline{1-3}
			& & \\[-10pt]
			$|S^\text{O}T^\text{I}\rangle$ &2 & $\left(|\uparrow_\text{R1}\downarrow_\text{L1}\rangle+|\uparrow_\text{L1}\downarrow_\text{R1}\rangle\right)\left(|\uparrow_\text{M2}\downarrow_\text{M3}\rangle-|\uparrow_\text{M3}\downarrow_\text{M2}\rangle\right)$ \\
			\cline{1-3}
			& & \\[-10pt]
			$|T^\text{O}S^\text{I}\rangle$ &2 & $\left(|\uparrow_\text{R1}\downarrow_\text{L1}\rangle-|\uparrow_\text{L1}\downarrow_\text{R1}\rangle\right)\left(|\uparrow_\text{M2}\downarrow_\text{M3}\rangle+|\uparrow_\text{M3}\downarrow_\text{M2}\rangle\right)$ \\
			\cline{1-3}
			& & \\[-10pt]
			$|T_{\pm}^\text{O}T_{\pm}^\text{I}\rangle$ &2 & $|\uparrow_\text{M2}\uparrow_\text{M3}\downarrow_\text{R1} \downarrow_\text{L1}\rangle-|\uparrow_\text{R1}\uparrow_\text{L1}\downarrow_\text{M2} \downarrow_\text{M3}\rangle$\\
			\cline{1-3}
			& & \\[-10pt]
			$|T^\text{O}T^\text{I}\rangle'$ &6 & $|\uparrow_\text{M2}\uparrow_\text{M3}\downarrow_\text{R1} \downarrow_\text{L1}\rangle+|\uparrow_\text{R1}\uparrow_\text{L1}\downarrow_\text{M2} \downarrow_\text{M3}\rangle+\left(|\uparrow_\text{R1}\downarrow_\text{L1}\rangle-|\uparrow_\text{L1}\downarrow_\text{R1}\rangle\right)\left(|\uparrow_\text{M2}\downarrow_\text{M3}\rangle-|\uparrow_\text{M3}\downarrow_\text{M2}\rangle\right)$\\
			\cline{1-3}
		\end{tabular}
		\caption{Eigenstates of the subspace formed by six electrons when four electrons in dot M form a triplet ground state.}
		\label{tab:tripletGroundEigenstates}
	\end{table*}
	
		\begin{figure}[t]
		\includegraphics[width=0.9\linewidth]{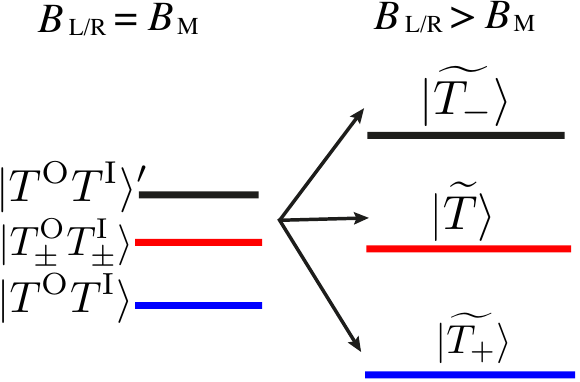}
		\caption{Schematic illustration of increased splittings between triplet states formed by outer electrons when applied magnetic fields on outer dots are stronger, $B_\text{L/R}>B_\text{M}$.}
		\label{fig:tripletSplitBLRgBM}
	\end{figure}

	Table \ref{tab:tripletGroundEigenstates} shows the six-electron eigenstates for the electron configuration shown as case (ii) in Fig.~\ref{fig:threecase}. The states are labeled using a Dirac ket with the first entry being the two-electron state of outer dots, i.e.~dots L and R, and the second entry being the four-electron state in dot M, denoted by the superscript ``O" and ``I" for outer and inner dots respectively. We have dropped the notations of core electrons in dot M for simplicity. For example, $\left|\uparrow_\text{M2}\uparrow_\text{M3}\downarrow_\text{R1} \downarrow_\text{L1}\uparrow_\text{M1}\downarrow_\text{M1}\right\rangle$ is written as $\left|\uparrow_\text{M2}\uparrow_\text{M3}\downarrow_\text{R1} \downarrow_\text{L1}\right\rangle$. As discussed in \cite{Martins.17,Malinowski.18,Deng.18}, the negative exchange energies for even number of electrons occupying a quantum-dot can be attributed to the magnetic correlations given by the ferromagnetic exchange term between two valence electrons \cite{Deng.18,Malinowski.18}, i.e. 
	\begin{equation}
		-J_\text{M2,M3}^F\left(\mathbf{S}_\text{M2}\cdot\mathbf{S}_\text{M3}\right),
	\end{equation}
	where $J_\text{M2,M3}^F>0$. The ferromagnetic exchange term reduces the energies of states with electrons in dot M forming triplets, $|T^\text{I}\rangle$, in relative to those forming singlets, $|S^\text{I}\rangle$. Excluding the high energy states with $\left|S^\text{I}\right\rangle$, we rewrite the remaining four lowest energy states in Table \ref{tab:tripletGroundEigenstates} as:
	$\left|\uparrow_{R_1}\downarrow_{L_1};T^\text{I}\right\rangle$, $\left|\uparrow_{L_1}\downarrow_{R_1};T^\text{I}\right\rangle$, $\left|\uparrow_{L_1}\uparrow_{R_1};T^\text{I}_{-}\right\rangle$, $\left|\downarrow_{L_1}\downarrow_{R_1};T^\text{I}_{+}\right\rangle$, where
	\begin{align}
		\begin{split}
			\left|T^\text{I}\right\rangle &= \left|\uparrow_{M_2}\downarrow_{M_3}\right\rangle-\left|\uparrow_{M_3}\downarrow_{M_2}\right\rangle, \\
			\left|T^\text{I}_{+}\right\rangle &= \left|\uparrow_{M_2}\uparrow_{M_3}\right\rangle, \\
			\left|T^\text{I}_{-}\right\rangle &= \left|\downarrow_{M_2}\downarrow_{M_3}\right\rangle.\\
		\end{split}
	\label{eq:tripletIntterStates}
	\end{align}

	In terms of singlet-triplet qubit whose exchange energy is mediated by electrons in dot M, we define:
	\begin{align}
		\begin{split}
			\left|\widetilde{S}\right\rangle &= \left|\uparrow_{R_1}\downarrow_{L_1};T^\text{I}\right\rangle + \left|\uparrow_{L_1}\downarrow_{R_1};T^\text{I}\right\rangle,\\
			\left|\widetilde{T}\right\rangle &= \left|\uparrow_{R_1}\downarrow_{L_1};T^\text{I}\right\rangle - \left|\uparrow_{L_1}\downarrow_{R_1};T^\text{I}\right\rangle,\\
			\left|\widetilde{T}_+\right\rangle &= \left|\uparrow_{R_1}\uparrow_{L_1};T^\text{I}_{-}\right\rangle,\\
			\left|\widetilde{T}_-\right\rangle &= \left|\downarrow_{R_1}\downarrow_{L_1};T^\text{I}_{+}\right\rangle,
		\end{split}
	\label{eq:logicalLeakageSt}
	\end{align} 
	where the former two states form the logical subspace and the latter two are the leakage states.
	When a uniform magnetic field is applied across the triple-quantum-dot device, the logical and leakage states are mixed, as shown in Table \ref{tab:tripletGroundEigenstates}. The mixture can be suppressed by applying different magnetic fields between the inner dot and outer dots, i.e. $|\Delta B| = |B_\text{L/R} - B_\text{M}| > 0$, cf. Fig.~\ref{fig:tripletSplitBLRgBM}, where $B_j$ is the magnetic field applied at dot $j$, as proposed for a pair of exchange-coupled singlet-triplet qubits \cite{Li.12,Wardrop.14,Cerfontaine.20.2,Buterakos.18}.
	
	\section{Leakage induced over $t_{2\pi}$} \label{sec:lea}
	\renewcommand{\theequation}{D-\arabic{equation}}
	\renewcommand{\thefigure}{D-\arabic{figure}}
	\setcounter{equation}{0}
	\setcounter{figure}{0}
	
	\begin{figure}[t]
		\includegraphics[width=0.9\columnwidth]{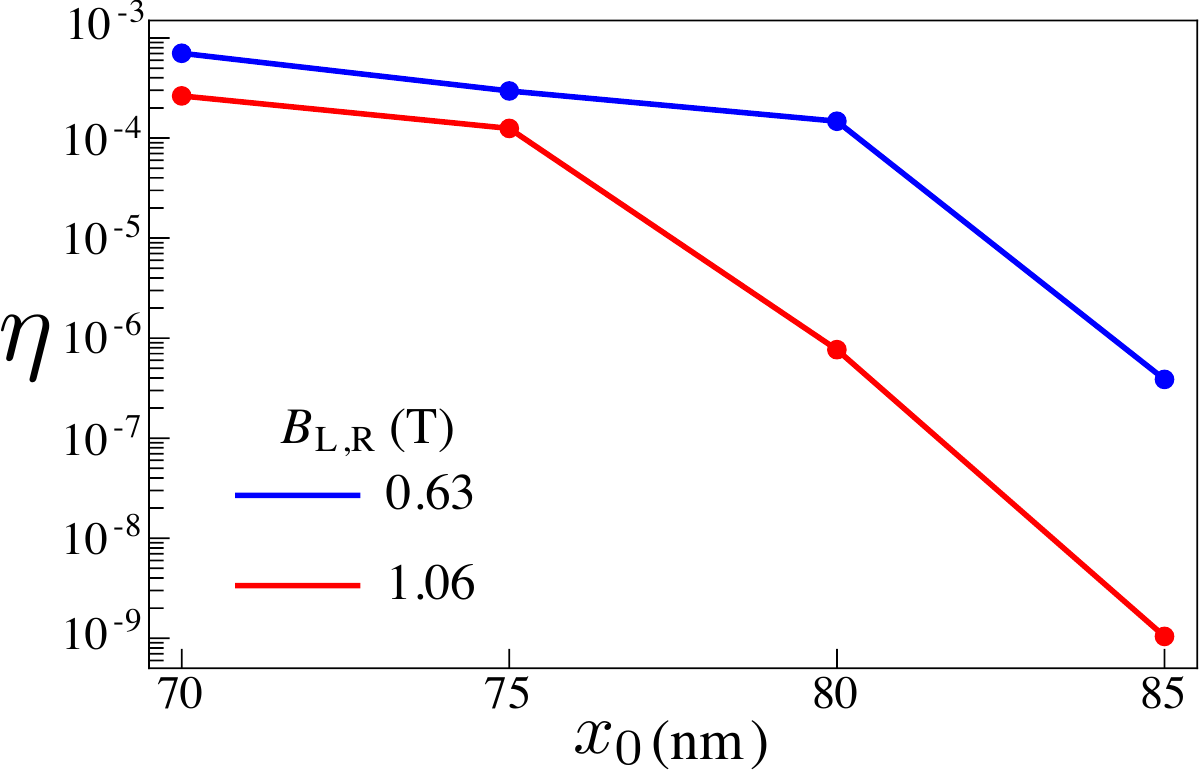}
		\caption{Leakage, $\eta$, as function of the magnetic field applied on outer dots, $B_\text{L/R}$. Magnetic field of strength $B=0.21 $T is applied at dot M.}
		\label{fig:leaVSBfoLR}
	\end{figure}

	In general, to suppress the leakage into $|\widetilde{T}_+\rangle$, $|\widetilde{T}_-\rangle$, larger $\Delta B$ is preferable. Also, we have found that the $\Delta B$ induced splittings between $|\widetilde{T}_+\rangle$, $|\widetilde{T}_-\rangle$ and $|\widetilde{T}\rangle$, is not straightforward when we take into account the orbital effects by magnetic field on the Fock-Darwin (F-D) states. To evaluate the leakage, using the CI results, we first identify the compositions of $|\widetilde{T}_+\rangle$, $|\widetilde{T}_-\rangle$ and $|\widetilde{T}\rangle$ in terms of Slater determinant and the corresponding orthonormalized F-D states. Then, we extract the tunneling energies between them by taking $t_{\widetilde{T},\widetilde{T}_\pm}=\langle \widetilde{T} |H|\widetilde{T}_\pm\rangle$. Leakage is estimated as:
	\begin{align}
		\begin{split}
			\eta &= \sum_{j\in \{+,-\} }\langle \widetilde{T}_j | \exp\bigg(-i \bigg[ \frac{J}{2} \left(|\widetilde{T}\rangle \langle \widetilde{T}|-|S^\text{O}T^\text{I}\rangle\langle S^\text{O}T^\text{I} |\right) \\
			&\quad+t_{\widetilde{T},\widetilde{T_j}} \left(|\widetilde{T}\rangle  \langle \widetilde{T_j} | +\text{H.c.}\right)\bigg]t_{2\pi}\bigg) |\widetilde{T}\rangle,
		\end{split}
	\end{align}
	where $t_{2\pi}=2\pi \hbar/J$ and $J=\langle \widetilde{T} |H|\widetilde{T}\rangle-\langle S^\text{O}T^\text{I} |H|S^\text{O}T^\text{I}\rangle$. The resultant leakage is shown in Fig.~\ref{fig:leaVSBfoLR}.

	\bibliographystyle{apsrev4-1}

\begin{thebibliography}{72}
\bibitem [{tri()}]{triplet}%
		\BibitemOpen
		\href@noop {} {}\bibinfo {note} {For simplicity, we will refer the
			unpolarized ($S_z=0$) triplet state as the ``triplet state'' in the remainder
			of the paper.}\BibitemShut {Stop}%
\end{thebibliography}
	%
	
\end{document}